# Improving pathology reports using business intelligence techniques: An experimental study

**Yasmine Alahmdai**
Decision Support Systems Laboratory
Monash University
Melbourne, Victoria
Email: yasminealahmadi@gmail.com

**Peter O'Donnell**
Decision Support Systems Laboratory
Monash University
Melbourne, Victoria
Email: peter.odonnell@monash.edu

## Abstract

Health professionals use pathology reports to monitor and manage a patient's health. Typically, pathologists diagnose patients' conditions and produce these reports which are then used as reference by clinicians and eventually shared with the patient. Pathology reports are difficult to interpret as the reports are written using complex medical terminology. As patients only see their doctors for a limited time, the complexity of report content and the manner in which the information is presented in the reports may hinder patients' understanding of their medical condition and prognosis. The objective of this study was to compare patient comprehension of results from two pathology-reporting styles: the traditional format in current widespread use and new style developed using techniques common in business intelligence system (BI) development. The study found that the reports prepared using a "BI style" improve experimental subject's understanding and satisfaction with the reports.

## 1 Introduction

This paper reports on a project that has investigated an approach to the improvement of the format of pathology reports. Pathology reports, which detail the results of medical tests, are a vital tool in the diagnosis and treatment of a wide range of medical conditions. The information is typically presented in a manner that pays little attention to the aesthetics or readability of the information the report contains. They typically resemble the kind of business report that was produced by old-style MIS reporting systems (circa the 1970s) on line printers. Almost universally a mono-spaced font is used, information is described with little supporting contextual information and the report data is labelled with technical shorthand notation. Some pathologists have expressed concern that the current report format that is currently used as a standard in the industry in both Australia and around the world may be contributing to poor communication of the pathology results. This may lead to doctors and patient misinterpreting the results and directly lead to poor health outcomes.

Standard business reports, of the kind produced by business intelligence (BI) systems, provide a style and format that, if applied to pathology reporting, could lead to a more effective format for the presentation of pathology reports. Through the application of some fundamental information presentation principles, foundations of business intelligence reporting, a new style of pathology report can be created. The study described in this paper investigates if reports created using these principles provide for better patient understanding of the results of pathology tests.

The paper is structured as follows: the first section discusses pathology reporting, describing current industry practice. The potential problems of current practice are discussed. A discussion of the manner in which the application of BI reporting principles to pathology reports could lead to improve reporting practice is then presented. The design of a web-based experiment follows. This includes a description experimental tasks and the hypothesis that they are used to test. The results of the study are then presented, discussing in turn, the demographics of the participants, the results of the tasks they were asked to perform, and the testing of the hypothesise. The paper then concludes, highlighting the findings of the study and the implications of the results for practice and for future research.

## 2   Pathology Reporting

In Australia, pathology services are provided to medical practitioners to enable "accurate diagnosis, management and prevention of disease. 70% of all medical diagnosis and 100% of all cancer diagnoses rely on a pathology report for diagnosis and care management" (Pathology Australia, 2015, p.1). Pathology represents a large and important segment of the healthcare economy. It has been estimated that pathology services in Australia generated over $2.8b revenue in 2014-2015 (IBIS World 2015). As the population ages and chronic conditions become more common, the sector is expected to steady grow at an annual rate of around 5% in the coming years (IBIS World 2015).

Medical practitioners order pathology tests to assess a patient's condition. Samples are taken and forwarded to a laboratory where they are examined and tested by pathologists. The pathologists then prepare a report that places the results in context of the patient's characteristics, the condition or disease involved, and the recommended approach for healthcare (Royal College of Pathologists of Australia, 2015). This report is then forwarded to the initiating medical practitioner for discussion in consultation with the patient leading to decision making about any required treatments or follow up tests. Typically, at the end of the consolation session, the patient will be provided with a printed copy of the report that they will take home with them.

Great care is taken when collecting and analysing samples to ensure errors are not made. In Australia, pathology laboratories are accredited by the National Association of Testing Authorities and the Royal College of Pathologists of Australasia for the Australian Government's Health Insurance Commission (Royal College of Pathologists of Australia, 2015). They apply policies similar to those used in the United States and the United Kingdom. Operations of Australian public and private laboratories are directed and assessed according to protocols of the International Standards Organisation and the National Pathology Accreditation Advisory Council, including reporting protocols (National Association of Testing Authorities, 2014). The main focus of these quality standards and protocol is the way samples are handled, tested an analysed. The reporting of results is governed by a standard that has only recently been created. A major focus of this standard developed by the Royal College of Pathologists of Australia (2013) has been the units and terminology used in reports. The standards are extensive and cover seven major areas of pathology with a decision-tree style mapping that identifies all factors of the patient's progress through the diagnosis and treatment of the pathology: anatomical, chemical, cytopathology, genetic, haematology, immunopathology and microbiology (Royal College of Pathologists of Australia, 2013).

In developing the standards, the College notes the use of the computer and Internet-based systems has led to wider dissemination of the reports, including various sources for the pathology results, and in need for pathology reports to be integrated with other health records. Results from the reports are also now frequently used in comparative displays and in computerised decision support in widely different healthcare settings including hospitals, community, indigenous health services and homes (Royal College of Pathologists of Australia, 2013).

The College has noted that the grammar, information structures and terminology used in reports are prone to misinterpretation, and this impacted records, subsequent analysis and decisions, and communication. The standards developed have addressed issues for computerisation of pathology reports including terminology, grammar and descriptive format; however, they did not address the physical and visual presentation of reports (either on paper or on screen). Mies argues that a pathology report should provide clear, unambiguous and complete diagnostic report to the medical practitioner who requested it, and that the report itself is "a permanent record of findings to guide patient care and ensure accountability" (Mies, 2015, p.185).

Aumann et al. (2013, p.387) argue that in a well designed pathology report as a result of "the uniform and clear layout of the report, the key findings can be recognized at a first glance." Almost universally, little attention is paid to the visual format and layout of pathology reports. Compared to the reports that are routinely used in business for routine tasks such as analysis of sales and costs, pathology reports look very old fashioned and would seem ineffective if their purpose is to convey – at a glance – the key findings of the pathology testing in an unambiguous manner. Figure 1 shows a typical pathology report. This report makes no use of the simple and generally accept principles of report design that are commonplace in reporting used in BI systems. For example, there is no use of any color coding to highlight key findings. The use of a non-proportional font also hinders readability.

```
Kidney Function                    Units      Ref Range

Sodium              142            mmol/L     135-145
Potassium           4.4            mmol/L     3.5-5.5
Chloride            80             mmol/L     95-110

Bicarbonate         28             mmol/L     20-32
Urea                5.8            mmol/L     3.0-8.0
Creatinine          85             umol/L     60-110
eGFR                >90
Anion Gap           18             mmol/L     10-20
Bilirubin           17             umol/L     4-20

Liver Function
ALP                 62             U/L        35-110
GTT              *  66             U/L        5-50
ALT                 20             U/L        5-40
AST                 27             U/L        10-40
Protein             75             g/L        66-83
Albumin             44             g/L        36-47
Globulin            31             g/L        26-41
```

*Figure 1: A typical pathology report. This report is formatted in the style that is the standard in use today. This particular report shows the results of a variety of tests of kidney and liver function.*

The format that is currently used to display pathology reports creates the possibility that key results might be misunderstood or simple missed by the medical practitioners when they read the reports. Further, it is even more likely that the results will not be understood by patients. Developing effective methods of communicating pathology results is an important factor in securing a patient's commitment to treatment, as well as in preventing misunderstanding of the significance of pathology report findings. In a review of the literature on pathology reports, Mossanen et al. (2014) noted that "not a single article addressed the patient as a stakeholder in the content of the pathology report. Nor did any article discuss the need for patient-centred pathology reports" (Mossanen et al. 2014, p.2192). Alzougool et al. (2013, p.12) states that "commonly used formats are often confusing to patients, and misunderstanding of reports can lead to negative outcomes for them."

## 3    Improving Pathology Reporting

The presentation of a pathology report may either assist comprehension or prove to be confusing if the primary findings are not clearly indicated (Mies, 2015). An effective design, according to Mies, (2015), should be able to "(1) communicate the pathologist's comprehensive analysis of facts, i.e., the diagnosis, and (2) create a permanent record of findings to guide treatment and ensure accountability" In addition, Valenstein (2008) also recommended that the report design should serve to improve the reader's recall and positive response, and enhance interpretation and comprehension. Mulsow et al. (2012) pointed out that these outcomes lead to potentially improved long term health outcomes.

Business Intelligence (BI) systems present data to business users in a wide variety of organisations and problem areas. Simple techniques are used to provide display of information that allow the key area of concern to be quickly identified and examined. For example, BI developers have routinely develop "dashboard" displays of key organisational data. "The information dashboard is a single screen display of the most important information people need to do a job, presented in a way to allow them to monitor (the data) in an instant (and) is a powerful new medium of communication." Few (2013, p.1).

These simple techniques and ideas – standard practice in BI systems - can be applied to the information presented in pathology reports, providing and improvement in the current visual design of the reports, enhancing the comprehension and interpretation of the content of the report.

These techniques common in BI system used could be used to present a range of data analyses either as static or dynamic representations of a pathology report, presenting colour-coded elements such as charts, figures, buttons, illustrations, pictures. These communication elements could be displayed successively to convey the situation to a patient, and allow him or her to explore the implications through the interface (Clark et al., 2013; Gaspar et al. 2013; Grant & Wheatley, 2014). This study aims to investigate the use of BI dashboard reporting-style to simplify interpretation of pathology results to enhance patient understanding and assist decision-making. Figure 2 shows an example of the way a pathology report could be formatted. This report shows the same pathology results – tests of kidney and liver function – that are depicted in Figure 1. This report has been developed used a proportional font, the data is formatted and groups into table using white-space, key results are highlighted using color, and icons are used to help lead the reader of the report to additional relevant explanatory text. To the developer of a BI system the report showing in Figure 2 would look like the kid of business report they create on a daily basis. In contrast, Figure 1 looks like a business report created using approaches and technologies of the 1970s.

## 4 The Research Design

### 4.1 Experimental Procedure

The main research question addressed but this paper is can pathology reporting be improved through the use of business intelligence report design techniques. A web-based experiment was designed and conducted to investigate this question. Subjects for the experiment were recruited by a call for participation placed on social media. The call for participation contained a link which directed interested people to a Web-based system used to provide information about, and to conduct, the experiment.

When first accessing the site potential subjects were greeted with a brief introduction to the experiment. If they were willing to participate, subjects acknowledged their consent by clicking a button which started the experimental procedure. Subjects were asked to responded to a simple set of questions used to collect basic demographic information. Once those answers were recorded the the system randomly allocate the subject to either the experimental group or the control group. Subjects did not know into which group they had been allocated. Subjects were then asked to examine and interpret three pathology different reports and answer questions about the content of the reports. They did this one report at a time, they were satisfied with their answers, participants clicked to submit their answers and to go to the next screen. Once they had completed those three tasks, they are shown another screen that redisplayed their answer to the first report (their description of the purpose of the report) they viewed. With just that prompt to help them remember the report, they were asked to recall and enter the the key results of the report. Next the subjects were asked their opinions of the reports they had viewed. Finally, subjects were taken to a closing screen which thanked them for their participation. They could leave an email address – not connected to the data collected from them – if they wanted to receive a summary of the results of the experiment.

### 4.2 The Analysis Tasks

The pathology reports shown to subjects in the experiment showed the results of different types of three common pathology tests. Only three reports were shown in order to avoid overwhelming study participants with too much medical information, and to keep the time required to complete the experimental process to a minimum. The reports included in the experiment showed the results of the following pathology tests (in the order they were shown to the subjects): tests of cholesterol levels, tests of kidney and liver function (see Figure 1 for the report shown to the control group, and Figure 2 for the report shown to the experimental group), and a set of blood tests. Each report was presented on a single Web-page along with with a brief scenario explaining the background of the patient for whom the ordered. This page allowed subjects were provided with the ability to "zoom" in on the report, though the report was clearly readable without being zoomed. On the same screen the questions were posed and a place to input their answers was provided. For each of the three reports subjects were asked to answer the same two questions: What is the purpose of the report? What are the key results? While the report style varied depending on whether the subject had been allocated to the experimental group (viewing "BI style" pathology reports) or the control group (viewing "traditional" style pathology reports).

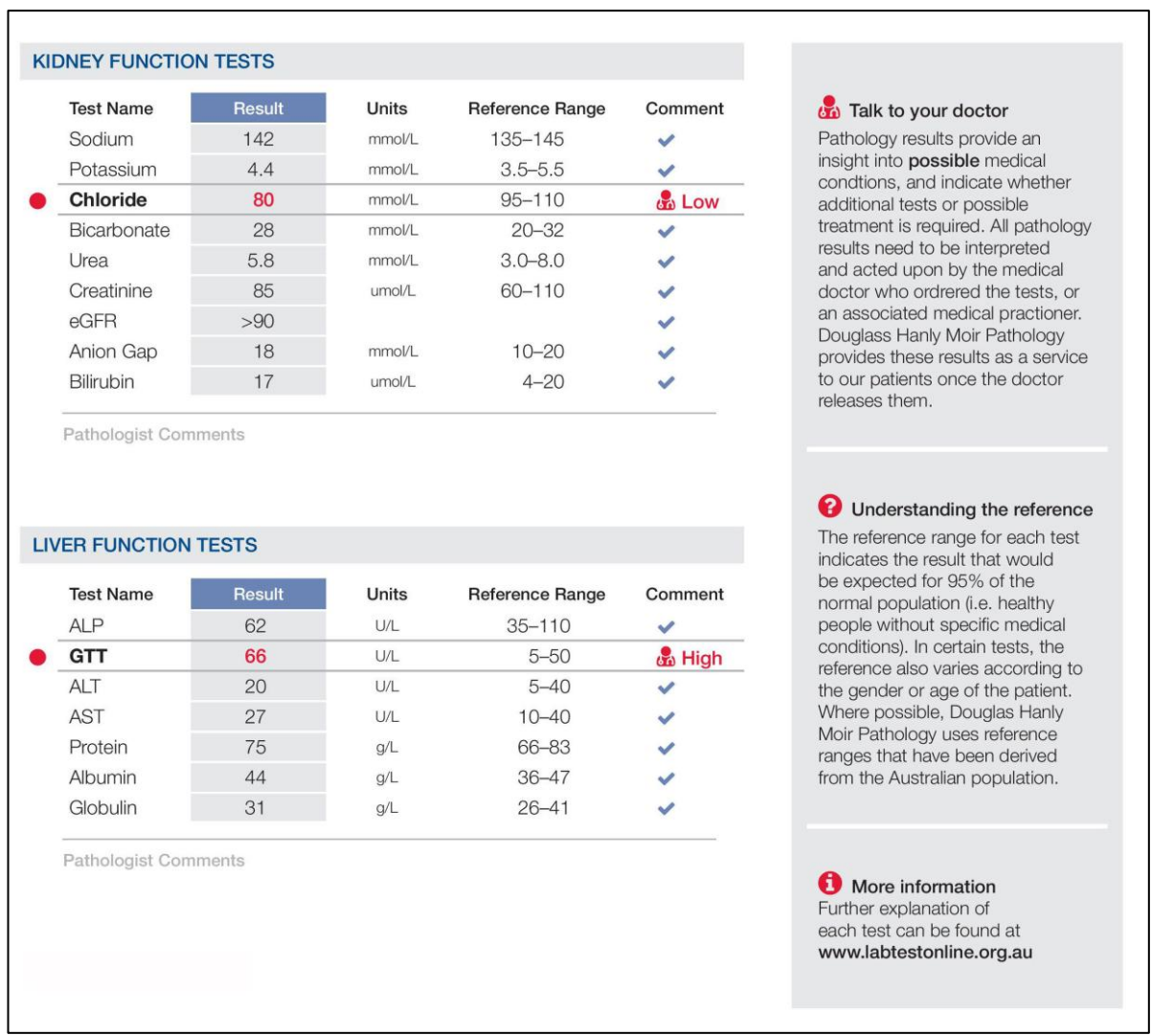

*Figure 2: A pathology report present in the "improved" style used in this study. This report shows the same results as the report depicted in Figure 1*

The pathology reports showed the results of different types of common pathology tests, the three reports showed the results of the following types of tests (in order): tests of cholesterol levels, tests of kidney and liver function (see Figure 1 for the report shown to the control group, and Figure 2 for the report shown to the experimental group), and a set of blood tests. Each report was provided with a brief case situation explaining the background of why the report was ordered, the report was shown – and subjects were provided with the ability to zoom in on the report – and on the same screen the questions were posed and a place to input their answers provided.

### 4.3  Measurement

When a subject began the experimental procedure their IP address was recorded and the time they started the first task was also recorded. (The IP addresses were recorded for quality control). At the end of the analysis tasks the time was again recorded allowing the total time for the analysis tasks to be determined. Where possible subject inputs were recorded using Web-based allowing them easily and unambiguously select the answer that they wanted. However, for two important questions subjects were asked to enter their responses via a text input field. The answers to the question "What is the purpose of the report?" was entered in this way for each of the three pathology reports viewed. The answers to the question "What are the key results?" was for the three pathology reports selected from a variety of options (in the form of a multi-choice question) – one of which was correct. When asked about the first report they had viewed – after viewing all three reports – subjects were promoted with their text answer to the question about the purpose of the report, and were again invited to answer the question "What are the key results?". This time they answered that question by entering test into a text entry field. These text based responses to the questions were coded as either correct or incorrect by the researchers after the data collection period had ended. In order to maintain consistency and to be as objective as possible a codebook was created to help guide the assessment of these items. This was used when coding subject's responses to these questions. When this coding was done, the coder was unaware of the group to which the subject belonged. All responses were coded by the same person, however, any ambiguous responses were checked with a second coder.

There are no survey instruments that have been designed to specifically measure patient satisfaction with the quality of pathology reports. In this study, the instrument designed by Doll and Torkzadeh's (1988) to measure end-user satisfaction with an information system – was adapted for this purpose. The Doll and Torkzadeh instrument contains questions in 5 groups (content, accuracy, format, ease to use and timeliness). In this study, two question groups weren't relevant and were removed form the instrument. The working of the remaining questions modified to suit the nature of the experiment.

The following 8 questions were used to measure the satisfaction subjects had with the content, format and ease of use of the reports they had viewed.

- If it were you, do you think the information content of the report meets your needs?
- Do you think, the reports you just viewed, provide sufficient information?
- Do you think that these reports communicate pathology information and key results effectively?
- Are you satisfied with the layout of the pathology reports?
- Do you think the output is presented in a useful format?
- Was the information clear?
- Is the design of the pathology reports that you just viewed user friendly?
- Are the reports that you just viewed easy to understand?

Subjects responded to each of these questions using a Web-based form with radio buttons that allowed them to indicate their response to the question using a bi-polar 6-point scale.

### 4.4 Hypotheses

Each of the hypotheses will compare the results obtained from the subjects who examined the "BI style" pathology reports (the experimental group) and those who examined the "traditional" style pathology reports (the control group). Hypothesis 1 examines the accuracy of the answers the subjects in the gave to the questions posed in the three experimental tasks. Hypothesis 2 examines the ability of subject to recall the information in the first report they examined after they had been distracted by another task. Hypothesis 3 examines a number of measures of satisfaction the subjects reported with the reports they examined. The hypotheses are summarised in Table 1.

| Hypothesis | Dependant variable(s) | Description |
| --- | --- | --- |
| 1 | Number of correct answers | Subjects viewing "BI style" pathology reports will get more correct answers to questions requiring understanding of the information shown than subjects who view "traditional" style reports. |
| 2 | Number of correct answers (after distraction) | Subjects viewing "BI style" pathology report will be more likely to recall information (once they have performed another task) than subjects who view "traditional" style pathology reports. |
| 3 | Subject responses to questions measuring perceived satisfaction. | Subjects viewing "BI style" pathology reports will respond more positively to the set of satisfaction measures than the subjects who view the "traditional" style pathology reports. |

*Table 1. A summary of the hypotheses tested.*

### 4.5 Data Treatment

The Web-based experimental system was created using the scripting language PHP. Results were stored directly in a MySQL database. Once the call for participation had been made on social media they system was open for subjects to use for a two-week period in June 2015. Once the data collection phase ended, the data was cleaned and prepared for the analysis stage by using Microsoft Excel 2013. The cleaning stage involved eliminating three duplicated records, deleting six multiple submissions data records. These submissions were identified by their identical IP addresses. Three out-of-range data items – errors in the recording of the time taken to compete the were also eliminated. Other processes such as replacing the empty values with Null keyword and decoding the categorical data to numerical values were performed. Coding of the answers of subject's free text responses to the question "What is the purpose of this report", and also of the question – in the memory recall task – "What were the key results of the report" were coded by the researchers using a code book (the process describe in section 4.4 Measurement). The data was then transferred to the STATA, a statistical software package, which was used to perform statistical analysis including the hypotheses testing that is presented in the next section of the paper (Results).

# 5 Results

## 5.1 Participant Demographics

Total of 154 participants completed the experiment, 82 participants were allocated to the experimental group and 72 participants to the control group.

Most of the participants (58 %) were aged from 25-34, while 24% belonged to the 35-44 age category. A similar portion of participants (9% and 8%) belonged to the age groups 18-24 and 45+ respectively. In regard to higher education levels, just over a third of the participants had a master's degree (37%) while 16% of the participants were at a PhD education level, and 10% reported that they had completed some postgraduate studies. 26% of the participants had a Bachelor's degree and approximately 8% had completed some college. A similar percentage of participants (29% and 25%) had worked in either an education, health or community development field, while 21% worked in information technology related business.

## 5.2 Results Summary

Table 2 shows the number of correct answers for the each of the experimental tasks for the subjects in both the experimental and control groups.

In regard to the questions "What is the purpose of the report?", "What are the key results?" and the memory recall question, the findings reveal that participants in the experimental group performed slightly better with respect to the majority of the questions. The total number of correct answers chosen by participants in the experimental group was 331 compared to 238 in the control group. In particular, for the liver and kidney function pathology report participants in the experimental group selected over twice the number of correct answers (51 and 63) compared to the participants in the control group.

|  |  | Number of Correct Answers | | % of Correct Answers | |
|---|---|---|---|---|---|
| **Report** | **Question** | **Experimental Group (n = 82)** | **Control Group (n = 72)** | **Experimental Group (n = 82)** | **Control Group (n = 72)** |
| Cholesterol tests | Report purpose | 49 | 36 | 59.7% | 50.0% |
|  | Key findings | 53 | 44 | 64.6 | 61.1 |
| Kidney & liver tests | Report purpose | 51 | 18 | 62.2 | 25.0 |
|  | Key findings | 63 | 37 | 76.8 | 51.3 |
| Blood tests | Report purpose | 35 | 43 | 42.7 | 59.7 |
|  | Key findings | 46 | 36 | 56.1 | 50.0 |
|  | **Total** | **331** | **238** | **67.3** | **55.1** |
| Recall task | Key findings | 35 | 24 | 42.7 | 29.2 |

*Table 2. A Summary of the key results of the report comprehension tests*

The subject's responses to the 8 questions recording aspects of their satisfaction with the reports. They rated their responses using a 6-point scale. "1" represented a strong negative response, "6" represented a strong positive response with a mid-point of 3.5.

## 5.3   Hypothesis Testing

To test hypothesis 1 and and 2, a two sample Wilcoxon Rank Sum (Mann-Whitney, 1947) test was used as the data for these questions is non-parametric.

Hypothesis 1 tests if the subjects who viewed "BI style" reports got more correct answers than the subjects who viewed "traditional" style reports. When ranking the number of correct responses, the rank sum of the experimental group was 7043, with and expected rank sum of 6355 (n=82), while for the control group the rank sum was 4082 with and expected rank sum of 5580 (n=72). That gives a Z score for the test of -2.525 with an associated P value of 0.0116. That means at a 95% confidence level the null hypothesis (that there is no difference between the groups) can be rejected and the alternative hypothesis can be accepted (that the subjects viewing "BI style" reports did outperform those viewing "traditional" reports.

| Question | Experimental group (n = 82) | | Control group (n=72) | |
|---|---|---|---|---|
| | Median | Mean | Median | Mean |
| If it were you, do you think the information content of the report meets your needs? | 5 | 3.83 | 4 | 2.95 |
| Do you think, the reports you just viewed, provide sufficient information? | 4 | 3.63 | 3 | 2.87 |
| Do you think that these reports communicate pathology information and key results effectively? | 4 | 3.36 | 3 | 3.14 |
| Are you satisfied with the layout of the pathology reports? | 5 | 3.69 | 3 | 3.17 |
| Do you think the output is presented in a useful format? | 5 | 3.50 | 3 | 3.08 |
| Was the information clear? | 4 | 3.41 | 3 | 2.99 |
| Is the design of the pathology reports that you just viewed user friendly? | 5 | 3.57 | 3 | 2.65 |
| Are the reports that you just viewed easy to understand? | 4.5 | 3.60 | 3 | 2.83 |

*Table 3. A summary of the results of the subject responses to the questions asking about their satisfaction with the reports they examined.*

Hypothesis 2 examines the performance of subjects in a memory recall task. They were asked the key results of a report they had previously viewed. The hypothesis tested is that the subjects viewing "BI style" pathology report will be more likely to recall information (once they have performed other tasks) than subjects who view "traditional" style pathology reports. The rank sum of the experimental group was 4962.5, with and expected rank sum of 6355 (n=82), while for the control group the rank sum was 4440 with and expected rank sum of 5580 (n=72). That gives a Z score for the test of -3.997 with an associated P value of 0.494. That means at a 95% confidence level the null hypothesis (that there is no difference between the groups) cannot be rejected and the alternative hypothesis is not supported. The subjects viewing "BI style" reports did not outperform those viewing "traditional" reports.

Each of the sets of answers for the question set used to test hypothesis 3 were be tested individually using a Chi-Squared test (see Table 3). Each of these questions examined an aspect of satisfaction with the pathology reports. The first question asked about the content. For that item the value of Pearson's Chi-Squared Statistic is 14.122 (df=5), with a p value of 0.015. For the next question, asking if the information was sufficient, the Pearson's Chi-Squared statistic is 18.51 (df=5) and the p value 0.0.002. The next question asks about the effectiveness of the reports, The Pearson's Chi-Squared statistic is 16.63 (df=5), and the p value 0.005.  The 4[th] question was concerned with the layout of the reports, the Pearson's Chi -Squared statistic is 21.93 (df=5), and the p value 0.001. The 5[th] question asked about the format of the reports. The Pearson's Chi -Squared statistic is 14.52 (df=5), and the p-value 0.13. For the 6[th] question, asking about the clarity of the reports, the Pearson's Chi-Squared statistic is 16.31 (df=5), and the p-value 0.006. For the 7[th] question, asking about the user friendliness of the reports, the Pearson's Chi-Squared statistic is 38.83 (df=5), and the p-value 0.000. For the final question, asking if the reports were easy to understand, the Pearson's Chi-Squared statistic is 18.79 (df=5), and the p-value 0.002. Taken individually, each of these provides a 95% confidence level that the experimental group were more satisfied with the reports they viewed compared to the control group. This allows the null hypothesis to be rejected and the alternative hypothesis to be accepted.

# 6　Conclusion

This paper reports the results of an experiment that aimed to see if pathology reports designed using techniques and principles commonly used in business intelligence reporting could improve the ability of people to read, and understand the information presented. Subjects in the study who viewed and examined "BI style" reports outperformed the subject who owes "traditional" style reports. Subjects who viewed the "BI style" reports were more satisfied with the information content, format and ease of use of the reports. While this study was limited in many regards, the subjects aren't stakeholders, had no personal interest in the results, and didn't have a doctor to consult with to explain the results to them – all key differences between the experience of the subjects in this study and real consumers of pathology reports. Even given those limitations, the study has shown that improvements to pathology reporting can be made by simple adopting the techniques (and technologies) that have been commonly used in business reporting for a long time.

While the Royal College of Pathologists has noted the need for the informed cooperation of patients and the need for physicians to emphasise patient's responsibility for patient well-being at any stage of care (Muslow *et al.*, 2012; National Pathology Accreditation Advisory Council, Australia, 2013). This can only occur if the patient understands the nature of the health condition and can contribute to stabilising or ameliorating its effects through healthcare interventions. Well designed pathology reports are a vital tool in achieving that understanding. The standards for pathology testing practices published by the College should be extended to include guidelines for the format and style of the presentation of the information in the report.

# 7　References

# Copyright